\documentclass[aps,prb,reprint,groupedaddress,amsfonts,amssymb,amsmath]{revtex4-2}

\usepackage{lineno,hyperref,amsmath,graphicx}

\begin{document}

\title{Critical Temperatures of Hard-Core Boson Model on Square Lattice within Bethe Approximation}

\author{E.~L.~Spevak, A.~S.~Moskvin, Y.~D.~Panov}
\email[]{yuri.panov@urfu.ru}
\affiliation{Ural Federal University, Ekaterinburg, 620002, Russia}

\begin{abstract}
The short-range correlations are considered for a two-dimensional hard-core boson model on square lattice within Bethe approximation for the clusters consisting of two and four sites. 
Explicit equations are derived for the critical temperatures of charge and superfluid ordering and their solutions are considered for various ratios of the charge-charge correlation parameter to the transfer integral. 
It is shown that taking into account short-range correlations for the temperatures of charge ordering results in the appearance of the critical concentration of bosons, which restricts the existence domain of the solutions of charge ordering type. 
In the case of superfluid ordering with the assumption of short-range correlations, the critical temperature is reduced up to zero values at half filling. 
A phase diagram of the hard-core boson model is constructed with the assumption of phase separation within Maxwell's construction and it is shown that taking into account short-range correlations within Bethe approximation quantitatively approximates the form of phase diagram to the results of the quantum Monte Carlo method. 
\end{abstract}

\keywords{hard-core bosons \sep phase diagram \sep the Bogolyubov inequality \sep the Bethe approximation}

\maketitle

\section{Introduction}

Lattice boson models are extensively employed for the description of the systems demonstrating nontrivial
phase diagram with competing or mixed states. 
The hard-core boson model was initially suggested for rationalization of the features of phase diagram of
superfluid $^4$He~\cite{Matsubara1956,Gersch1963} 
and various variants of this model are currently used for the description of the characteristics of a large number of physical systems including 
high-temperature superconductors~\cite{Micnas1990,Lindner2010}, 
ultracold atoms in optical lattices~\cite{Dutta2015}, 
excitons in nanotubes~\cite{Abramavicius2012}, 
and magnetic insulators in external field~\cite{Giamarchi2008}. 
An interest in these models has grown in recent years after discovery of the coexistence of charge ordering and superconductivity in cuprates~\cite{Chang2012,Ghiringhelli2012,DaSilvaNeto2014}.

Hamiltonian of hard-core charged bosons is following~\cite{Matsubara1956,Micnas1990}: 
\begin{equation}
	\mathcal{H} =
	- t \sum\limits_{\left\langle i,j \right\rangle}  
	\left( b_{i}^{+} b_{j}^{} + b_{j}^{+} b_{i}^{} \right) 
	 + V \sum\limits_{\left\langle i,j \right\rangle}  n_{i} n_{j}  
	- \mu \sum\limits_{i} n_{i} ,   
	\label{eq:hcB}
\end{equation}
where $b^{+}\,(b)$ correspond to the creation (annihilation) operators of hard-core bosons with Bose-type commutation relations for different sites $[b_{i},b_{j}^{+}] = 0,$ $i\neq j$, 
and Fermi-type relations on one site: $\left\{ b_{i},b_{i}^{+} \right\} = 1$, 
$n_i = b_{i}^{+} b_{i}$ is the operator of the number of hard-core bosons on a site, 
$t$ and $V$ correspond to the transfer integral and parameter of charge correlations between nearest neighbors, 
and $\mu $ is chemical potential, which is necessary to take into account the condition of a constant concentration of bosons: $nN= \sum\limits_{i}\langle n_{i}\rangle$, where $N$ is the total number of sites. 
We further consider hard-core bosons on a planar square lattice.

\begin{figure*}
   \includegraphics[width=1.0\textwidth]{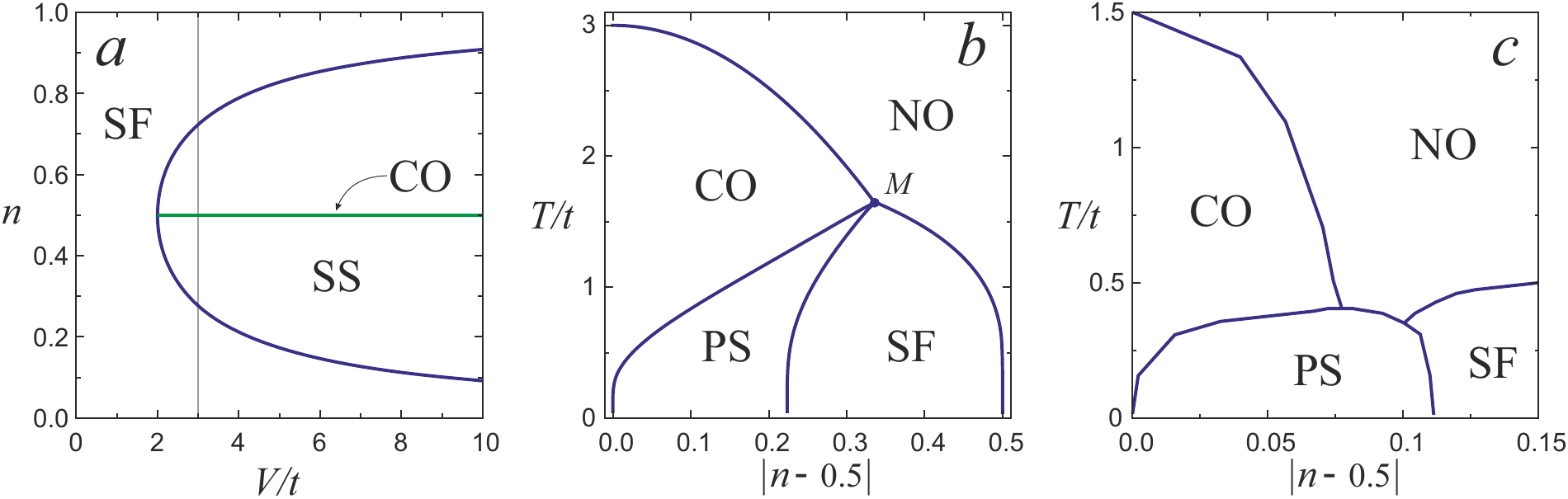} 
   \caption{
	(a) Diagram of the main state of model~\eqref{eq:hcB} to mean field approximation. 
	SF is the superfluid, SS is supersolid, and CO is charge ordering. 
	(b) Phase diagram within mean field approximation at $V/t = 3$ 
	(corresponds to gray vertical line on panel (a)). 
	PS is phase separation into macroscopic domains of SF and CO phases and NO is the non-ordered phase. 
	(c) Phase diagram obtained by quantum Monte Carlo at $V/t = 3$ 
	(according to the data from~\cite{Schmid2002}).
	}
\label{fig:Fig1}
\end{figure*}

The ground state diagram of the hard-core bosons within mean-field approximation (MFA)
is well known~\cite{Micnas1990,Robaszkiewicz1981} and is shown in Fig.~\ref{fig:Fig1}a in $(n,V/t)$ variables.
Superfluid (SF) represents the ground state at all $n$ values 
on the left from the Heisenberg point $V/t=2$; 
charge ordering (CO) is realized on the right at $n=1/2$; 
SF phase is realized at $|n-1/2|>\eta^{*}$, $\eta^{*}=\tfrac{1}{2}\sqrt{(V/t-2)/(V/t+2)}$; 
and the ground state is represented by supersolid (SS) at $0<|n-1/2|<\eta^{*}$, 
the phase, in which the charge and superfluid order parameters are non-zero.
The conditions for the existence of the SS phase and its stability against phase separation (PS) into macroscopic domains of SF and CO phases remain the subject of discussion and detailed studies. 
Numerical simulations using quantum Monte Carlo method show the stability of the SS phase in the models with frustrating interactions on triangular lattice~\cite{Wessel2005,Heidarian2005,Melko2005,Suzuki2013} 
or on square lattice with the interaction 
of the next-nearest neighbors~\cite{Dang2008,Ng2010,Capogrosso-Sansone2010,Kar2016}. 
A stable SS phase was also identified in various variants of the Bose--Hubbard model~\cite{Dutta2015}. 
PS and SS phases coincide in the case of model~\eqref{eq:hcB} within MFA in the main energy state and the existence domain, 
while the SS phase is metastable at a finite temperature~\cite{Batrouni2000,Hebert2001}: 
its energy is always larger than that of the PS state.
Phase $(n,T)$ diagram of model (1) within MFA [11] is given in Fig.~\ref{fig:Fig1}b for $V/t = 3$. 
Its interesting feature is the presence of the CO phase at $n\neq1/2$,
which transforms into PS state with a decrease in temperature.
The PS domain starts from a tricritical point $M$. 
Phase diagram of model~\eqref{eq:hcB} for $V/t = 3$ (Fig.~\ref{fig:Fig1}c) 
obtained in~\cite{Schmid2002} using quantum Monte Carlo method is completely different from the results of MFA; the scale of critical temperatures is much lower and the concentration range corresponding to the CO phase is narrower; in addition, the concentration range, at which a transition from the non-ordered (NO) phase to the PS phase occurs, arises instead of tricritical point.

A large number of studies of the lattice boson models is devoted to the calculations using quantum Monte Carlo. 
Analytically different characteristics of the hard-core boson system were studied using spin wave method~\cite{Micnas1990,Murthy1997,Pich1997,Bernardet2002,Altman2003} and different variants
of Green function method~\cite{Micnas1990,Micnas2007,Antsygina2009}, in addition to the mean-field method. 
In this work, effect of the short-range correlations within Bethe approximation was considered. 
In the case of the clusters consisting of two and four sites, 
critical temperatures of model~\eqref{eq:hcB} were found and evolution
of phase diagrams in $(n,T)$ variables depending on these approximations was considered.

The paper is organized as follows. 
Main equations for Bethe approximation in the hard-core boson model on a two-dimensional square lattice are written in section 2. 
Section 3 is devoted to the study of the behavior of critical temperatures for various $V/t$ ratios
and phase $(n, T)$ diagrams were constructed with the assumption of phase separation. 
We compare these results with MFA and quantum Monte Carlo. 
The last section is devoted to conclusions.

\section{Bethe approximation for clusters consisting of two and four sites}

The hard-core boson system is equivalent to quantum magnetic with a constant magnetization, the system of pseudospins $s=1/2$ in external field directed along the $z$ axis~\cite{Matsubara1956}, which is described by following Hamiltonian:
\begin{multline}
	\mathcal{H}
	= -t \sum_{\langle i,j \rangle}  \left( s_{i}^{+} s_{j}^{-} + s_{i}^{-} s_{j}^{+} \right)
	+ V \sum_{\langle i,j \rangle} s_{i}^{z} s_{j}^{z} \\
	- h \sum_{i} s_{i}^{z} 
	+ E_0
	,
\label{eq:spinH}
\end{multline}
where 
$s_i^{-}= b_i^{+} $, $s_i^{+}= b_i^{}$, $s_i^{z}=\tfrac{1}{2}-n_i$, 
$h = 2V - \mu$, and $E_0 =  N\varepsilon_0 = N ( V - \mu )/2$.

\begin{figure}
   \includegraphics[width=0.45\textwidth]{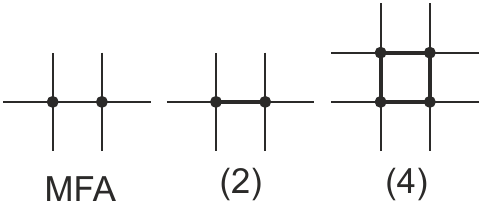} 
   \caption{
	The form of Bethe clusters used for calculation of Hamiltonian~\eqref{eq:Hc}. 
	MFA corresponds to the cluster from non-interacting sites. 
	Dark circles indicate the sites of sublattice $A$ and light circles indicate the sites of sublattice $B$. 
	}
\label{fig:Fig2}
\end{figure}

In order to evaluate large potential $\Omega(\mathcal{H})$, we employ Bogolyubov inequality: 
$\Omega(\mathcal{H}) \leq \Omega = \Omega(\mathcal{H}_0) + \left\langle \mathcal{H}-\mathcal{H}_0\right\rangle$, 
where zero-order Hamiltonian $\mathcal{H}_0$ appears as the sum of Hamiltonians $\mathcal{H}_c$ for non-interacting
Bethe clusters from $c$ sites covering the entire lattice:
$\mathcal{H}_0 = \sum \mathcal{H}_c$. 
Variants of the clusters employed by us are given in Fig.~\ref{fig:Fig2}. 
Let us define two sublattices $\alpha=A,B$ on a square lattice and write equation for $\mathcal{H}_c$ as follows:
\begin{multline}
	\mathcal{H}_c
	= -t \sum_{\langle i,j \rangle}^{c}  \left( s_{i}^{+} s_{j}^{-} + s_{i}^{-} s_{j}^{+} \right)
	+ V \sum_{\langle i,j \rangle}^{c} s_{i}^{z} s_{j}^{z} \\
	-  {\bf g}_f \sum_{i=1}^{c} {\bf s}_i 
	-  {\bf g}_a \sum_{i=1}^{c} (-1)^{\alpha_i} {\bf s}_i .
	\label{eq:Hc}
\end{multline}
Here, summation is performed over cluster sites, 
$(-1)^A=1$ and $(-1)^B=-1$, 
and ${\bf g}_{f,a}$ are molecular fields, which represent variational parameters and account for interaction of Bethe cluster with environment. 
MFA formally corresponds to Hamiltonian~\eqref{eq:Hc} containing only the terms with molecular fields.

Isotropic character of transfer in  $(x,y)$ plane allows one to restrict only to $x$ and $z$ components of molecular fields. 
After calculating the partition function for the cluster $Z_c=\mathop{\rm Tr \exp(-\beta\mathcal{H}_c)}$, 
where $\beta=1/k_B T$ (we further consider that $k_B=1$), 
we write equations in terms of molecular fields for the deviation of the concentration of bosons from half filling, 
$\eta = 1/2-n$, and various ordering parameters (OPs):
\begin{equation}
\begin{array}{ll}
	\eta = \dfrac{1}{c\beta} \dfrac{\partial \ln Z_c}{\partial g_f^z} ,&\quad
	\xi = \dfrac{1}{c\beta} \dfrac{\partial \ln Z_c}{\partial g_a^z} ,\\[1.5em]
	\rho = \dfrac{1}{c\beta} \dfrac{\partial \ln Z_c}{\partial g_f^x} ,&\quad
	\chi = \dfrac{1}{c\beta} \dfrac{\partial \ln Z_c}{\partial g_a^x} .
\end{array}
	\label{eq:op}
\end{equation}
Here, $\xi$ and $\rho$ correspond to charge and superfluid OPs and $\chi$ is the asymmetry of superfluid OP.

An estimation of the grand potential per site, $\omega=\Omega/N$, is following:
\begin{multline}
	\omega = -\frac{1}{c\beta} \ln Z_c + aV \left( \eta^2 - \xi^2 \right) - 2at \left( \rho^2 - \chi^2 \right) + \\
	+ \eta g_f^z + \xi g_a^z + \rho g_f^x + \chi g_a^x - h \eta + \varepsilon_0 ,
\end{multline}
where $a=2$ for MFA, $a=3/2$ for the cluster of two sites, and $a = 1$ for the cluster of four sites. 
Minimizing $\omega$ by molecular fields, we have the equations for OP:
\begin{equation}
\left\{
\begin{array}{l}
	2aV\eta = -g_f^z + h ,\\[1em]
	2aV\xi = g_a^z ,\\[1em]
	4at\rho = g_f^x ,\\[1em]
	4at\chi = -g_a^x .
\end{array}
	\right.
	\label{eq:opeq}
\end{equation}

We can find critical temperatures from the condition of loss of stability of minimum $\omega$ for the NO phase relative to the variation by corresponding molecular field. 
Following equation is derived for the temperature of charge ordering, $T_{CO}$:
\begin{equation}
	2aV \left.\left(\frac{\partial\xi}{\partial g_a^z}\right)\right|_0 = 1 ,
	\label{eq:Tco}
\end{equation}
whereas the equation is written as follows for the temperature of superfluid transition, $T_{SF}$:
\begin{equation}
	4at \left.\left(\frac{\partial\rho}{\partial g_f^x}\right)\right|_0 = 1 .
	\label{eq:Tsf}
\end{equation}
In Eqs.~\eqref{eq:Tco} and~\eqref{eq:Tsf}, derivative is calculated at $g_a^z=0$, $g_f^x=0$, and $g_a^x=0$. 
In order to obtain the dependence on the concentration of bosons, these equations need to be solved along with Eq.~\eqref{eq:op} for $\eta$. 
It should also be noted that analogous equation for critical temperature of disorder-order transition 
related to variation of $g_a^x$ have no solution.

The PS domain on the phase diagram in $(n, T)$ variables can be determined 
using Maxwell's construction~\cite{Kapcia2012,Kapcia2014,Kapcia2015}. 
Solving Eq.~\eqref{eq:opeq} individually for the CO phase and for the SF phase, we determine the parameter $\mu^{*}$, 
at which the values of grand potential are identical in these phases, 
$\omega_{CO}(\mu^{*})=\omega_{SF}(\mu^{*})$, and we determine boundary concentrations of the PS domain 
from Eq.~\eqref{eq:op} for $\eta$, $\eta_{CO}(\mu^{*})$ and $\eta_{SF}(\mu^{*})$.


\section{Results}

Eqs.~\eqref{eq:Tco} and~\eqref{eq:Tsf} within MFA result in well-known equations~\cite{Robaszkiewicz1981} for critical temperatures:
\begin{equation}
	\frac{T_{CO}}{V} = 1 - 4\eta^2 , \qquad
	\frac{T_{SF}}{2t} = 4\eta  \left( \ln \frac{1 + 2\eta}{1 - 2\eta} \right)^{-1} . 
	\label{eq:T1}
\end{equation}

In the case of the Bethe cluster of two sites, Eq.~\eqref{eq:Tco} for $T_{CO}$ has the following form:
\begin{equation}
	\frac{ \cosh(\beta t) + 2g }{ 2\sinh ( \beta t )}
	= \frac{3V}{4t} \left(1 - 4\eta^2\right) ,
	\label{eq:T2co}
\end{equation}
where
\begin{equation}
	g = \tfrac{1}{2}\sqrt{ \left(1 - 4\eta^2\right) e^{-\beta V} + 4\eta^2 \cosh^2(\beta t)}.
\end{equation}
Eq.~\eqref{eq:Tsf} for $T_{SF}$ taking into account the expression~\eqref{eq:op} for the Bethe cluster of two sites can be reduced to the following form:
\begin{multline}
	\frac{ x^2 - \left(V/2 + t\right)^2 }{6 t }=\\
	\left(V/2 + t\right) 
	\frac{2\eta^2 \cosh(\beta t) - 2e^{\beta t} \left(\frac{1}{4}-\eta^2\right) + g}{ \cosh(\beta t) + 2g}
	+ \eta x
	,
	\label{eq:T2sf}
\end{multline}
where 
\begin{equation}
	x = \beta^{-1} \ln\left( \frac{2\eta \cosh(\beta t) + 2g}{1 - 2\eta} \right) + V/2 . 
\end{equation}

\begin{figure*}
   \includegraphics[width=1.0\textwidth]{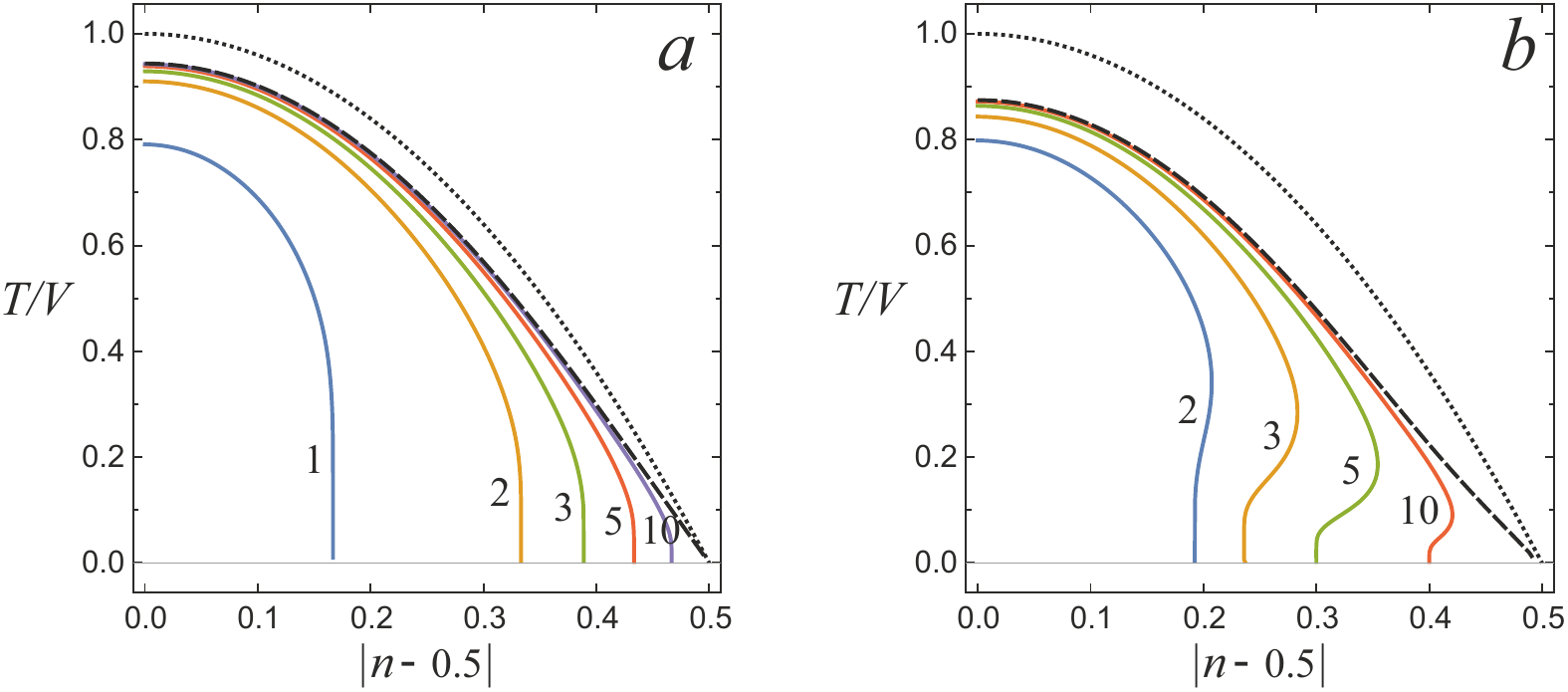} 
   \caption{
	Reduced critical temperatures $T_{CO}$ for 
	(a) two-site Bethe approximation, Eq.~\eqref{eq:T2co}, and 
	(b) four-site Bethe approximation, Eq.~\eqref{eq:T4co}. 
	The numbers near curves correspond to $V/t$ values. 
	Dashed line shows the limiting case, $V/t\rightarrow\infty$, 
	and dotted line shows the reduced critical temperature of charge ordering, Eq.~\eqref{eq:T1}.
	}
   \label{fig:Fig3}
\end{figure*}

In the case of the Bethe cluster of four sites, the explicit form of equations for critical temperatures becomes more cumbersome. 
Eq.~\eqref{eq:Tco} for  $T_{CO}$ can be written as follows:
\begin{multline}
	Z_0^{-1} 
	\bigg[ 
	\bigg( 2\beta V + \frac{V}{t}\sinh(2\beta t) \bigg) \cosh(\beta y) \\
	+ e^{\frac{1}{2}\beta V} \frac{V^2}{2t^2} \bigg( \cosh(\beta d) 
	+ \frac{V^2/2 + 8t^2}{Vd} \sinh(\beta d) \\
	- e^{\frac{1}{2}\beta V}\bigg) 
	\bigg] = 1 ,
	\label{eq:T4co}
\end{multline}
where $d=\sqrt{V^2/4 + 8t^2}$, 
\begin{multline}
	Z_0 = 3 + e^{\beta V} + 2 e^{\frac{1}{2}\beta V} \cosh(\beta d) \\[0.5em]
	+ 2 e^{-\beta V} \cosh(2\beta y) 
	+ 4 \left(1 + \cosh(2\beta t)\right) \cosh(\beta y) .
\end{multline}
Here, $Z_0$ means a partion function of Bethe cluster in the NO phase. 
Eq.~\eqref{eq:T4co} needs to be solved along with Eq.~\eqref{eq:op} for $\eta$, which can be written as follows in this case:
\begin{multline}
	Z_0^{-1} \sinh(\beta y) \Big(1 + \cosh(2\beta t) \\
	+ 2 e^{-\beta V} \cosh(\beta y)\Big) = \eta .
	\label{eq:eta4}
\end{multline}

Finally, Eq.~\eqref{eq:Tsf} for $T_{SF}$ can be written in explicit form as follows:
\begin{multline}
	2tZ_0^{-1}
	\bigg[ \:
	\frac{2}{y}\sinh(\beta y) \\[0.5em]
	- a e^{\frac{1}{2}\beta V} \left(b_1 \cosh(\beta d) + b_2 \sinh(\beta d)\right) + c e^{\beta V} \\[0.5em]
	- 2 e^{-\beta V} \left( c \cosh(2\beta y) + s \sinh(2\beta y) \right) \\[0.5em]
	+ e^{2\beta t} \left( b_3 c \cosh(\beta y) + b_4 s \sinh(\beta y) \right) \\[0.5em]
	- e^{-2\beta t} \left(c \cosh(\beta y) + s \sinh(\beta y) \right)
	\bigg] = 1 .\qquad
	\label{eq:T4sf}
\end{multline}
Here, the parameters 
\begin{equation}
	a = 2t+V , \quad
	c = a/(a^2-y^2) , \quad
	s = y/(a^2-y^2) , 
\end{equation}
and 
\begin{equation}
	\begin{array}{lcl}
		b_0 &=& y^4 - y^2 \left(24 t^2-4 t V+V^2\right) + 4 a^2 t^2 , \\[1em]
		b_1 &=& \left( y^2+4 t (V-7 t) \right)/b_0 , \\[1em]
		b_2 &=& \left( y^2 (8 t+V)- 4 t \left(40 t^2-5 t V+V^2\right) \right)/b_0 , \\[1em]
		b_3 &=& \left( y^4 -4 a^2 t (5 t-V) - y^2 (V-4 t)^2 \right)/b_0 , \\[1em]
		b_4 &=& \big( y^2 \left(32 t^2+4 t V+3 V^2\right) \\[0.5em]
		&&{}- 2 a^2 \left(30 t^2-8 t V+V^2\right) - y^4 \big)/\left(2d\,b_0\right) .
	\end{array}
\end{equation}
were introduced. Eq.~\eqref{eq:T4sf} should also be solved along with Eq.~\eqref{eq:eta4}.

Figure~\ref{fig:Fig3} shows solutions of Eqs.~\eqref{eq:T2co} and~\eqref{eq:T4co} for various $V/t$ ratios. 
In the case of MFA, the reduced critical temperature of charge ordering, $T_{CO}/V$, is described by a universal parabolic dependence on the concentration of bosons at all values of $t$ parameter according to Eq.~\eqref{eq:T1}. 
Accounting for the short-range correlations within the Bethe approximation results in a quantitative decrease in the $T_{CO}/V$ value and a fundamental change of the form of dependences of critical temperature of charge ordering on the concentration. 
For a finite value of the ratio $V/t$ a critical concentration of bosons arises, 
$|n_c - 1/2| = \eta_c$, which restricts the existence domain of critical temperature of charge ordering. 
In the case of two-site Bethe approximation, analysis of Eq.~\eqref{eq:T2co} 
gives the value $\eta_c = 1/2 - t/3V$, which could provide the critical value of $V/t = 2/3$, 
below which there is no solution of Eq.~\eqref{eq:T2co} for $T_{CO}$. 
It should also be noted that solutions of Eq.~\eqref{eq:T2co} are related to metastable states (for local minima of large potential) at the $V/t$ values below the Heisenberg point, $V/t<2$; 
whereas at $V/t>2$, solutions of Eq.~\eqref{eq:T2co} correspond to the NO--CO transition only up to the tricritical point. 
Numerical analysis of Eq.~\eqref{eq:T4co} shows that there are no solutions at $V/t < 1$ in the case of a four-site Bethe approximation. 
As follows from Fig.~\ref{fig:Fig3}, the limiting dependence for $T_{CO}$ at $V/t \rightarrow \infty$ exists in the entire concentration range; however, the maximum value of reduced temperature at half filling is less than the MFA value by 5.5\% in the case of two-site and by 12.5\% in the case of four-site Bethe approximation.

\begin{figure*}
   \includegraphics[width=1.0\textwidth]{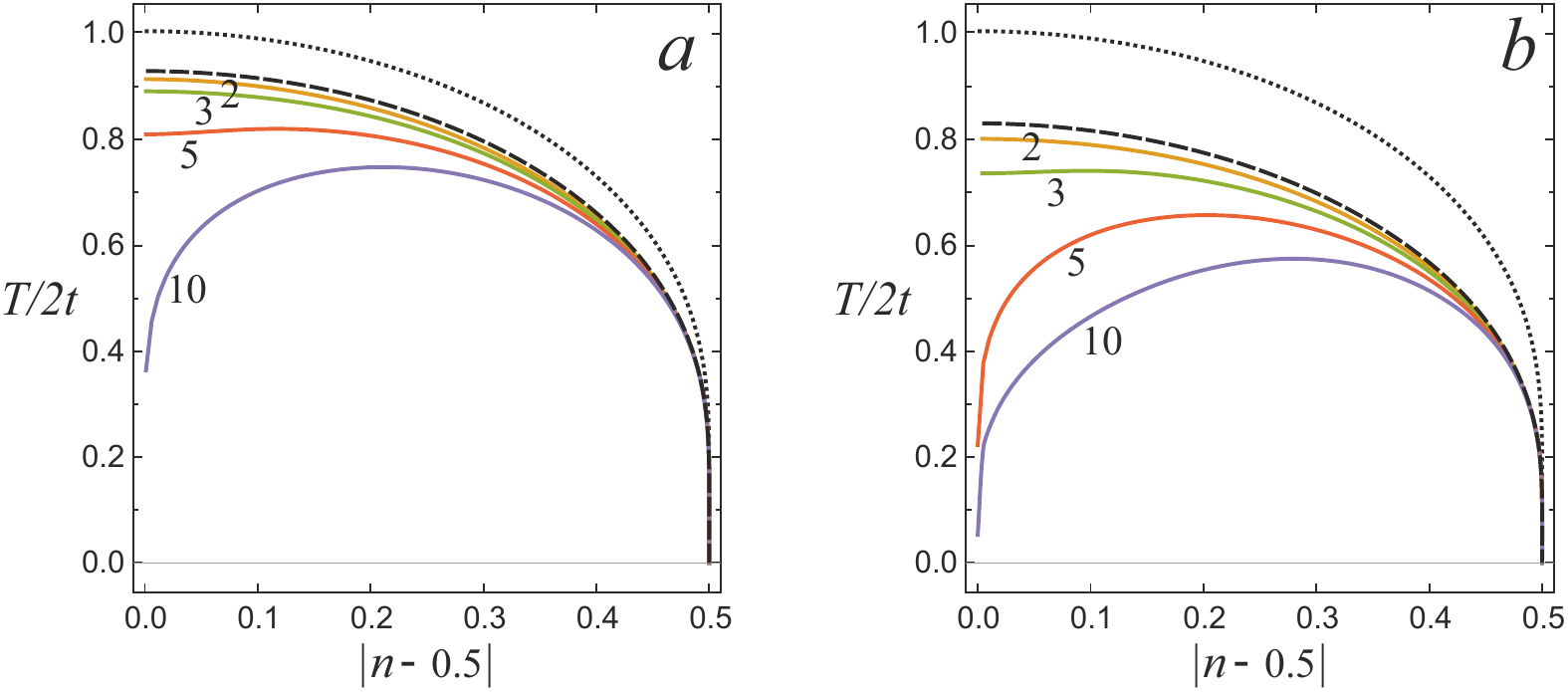} 
   \caption{
	Reduced critical temperatures $T_{SF}$ for 
	(a) two-site Bethe approximation, Eq.~\eqref{eq:T2sf}, and 
	(b) four-site Bethe approximation, Eq.~\eqref{eq:T4sf}. 
	The numbers near curves correspond to $V/t$ values. 
	Dashed line shows the limiting case, $V/t = 0$, 
	and dotted line shows the reduced critical temperature of superfluid ordering, Eq.~\eqref{eq:T1}.
	}
   \label{fig:Fig4}
\end{figure*}

The reduced critical temperatures of superfluid ordering, $T_{SF}/2t$, which represent solutions of Eqs.~\eqref{eq:T2sf} and~\eqref{eq:T4sf} at different $V/t$ ratios, are given in Fig.~\ref{fig:Fig4}. 
A universal dependence for MFA~\eqref{eq:T1} corresponds to the dotted line. 
Assumption of short-range correlations within two- and four-site Bethe approximation results in the reduction of $T_{SF}/2t$ value with an increase in the $V/t$ ratio. 
This effect is always more pronounced in the case of four-site approximation.
Fundamental change involves a decrease in $T_{SF}$ up to zero values at half filling with an increase in the $V/t$ ratio. 
The limiting dependence of the reduced critical temperature at $V/t \rightarrow 0$ possesses the maximum temperature at half filling as in the case of MFA; 
however, the magnitude is less than the MFA value by 7.5\% in the case of two-site and by 16.5\% in the case of four-site Bethe approximation. 
In contrast to charge ordering, solutions of Eqs.~\eqref{eq:T2sf} and~\eqref{eq:T4sf} for $T_{SF}$ exist in the entire concentration range $0<|n|<1/2$; however, a NO--SF transition corresponds only to the concentrations on the right from tricritical point at $V/t > 2$.

\begin{figure}
   \includegraphics[width=0.45\textwidth]{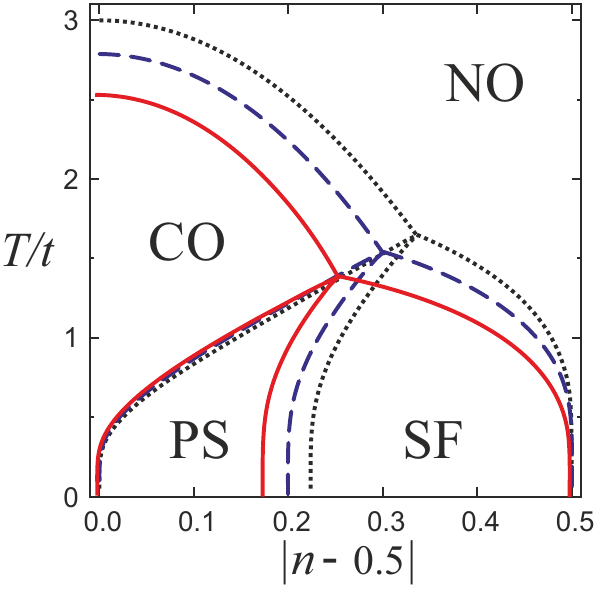} 
   \caption{
	Phase diagram of the hard-core boson model at $V/t = 3$. 
	Dotted line corresponds to MFA, dashed line corresponds to two-sitee Bethe approximation, 
	and bold line corresponds to four-site Bethe approximation. 
	The existence domains of non-ordered phase (NO), 
	charge-ordered phase (CO), 
	superfluid phase (SF) phase, 
	and phase separation (PS) are shown.
	}
   \label{fig:Fig5}
\end{figure}

Figure~\ref{fig:Fig5} shows the phase diagram of the hard-core boson model obtained within two- and four-site Bethe approximation at $V/t = 3$. 
Dotted lines indicate the results of MFA for comparison~\cite{Robaszkiewicz1981}. 
Taking into account short-range correlations leads to the narrowing of the domains for CO and PS phases; both maximum temperature and the concentration range decrease.
The maximum of the temperature of charge ordering at half filling decreases by nearly 7\% for two-site approximation and by 16\% for four-site approximation relative to MFA. 
In the case of quantum Monte Carlo simulations, such decrease corresponds to 50\%. 
The maximum temperature of ordering in the case of the SF phase also decreases after taking into account short-range correlations and the concentration range increases. 
Tricritical point shifts towards half filling along the CO--PS line. 
On the whole, assumption of neighboring correlations within two- and four-site Bethe approximation approaches the form of phase diagram to the results obtained by quantum Monte Carlo~\cite{Schmid2002}; 
however, fundamental feature related to the NO--PS transition is not reproduced within Bethe approximation; tricritical point represents the boundary point of the PS range within Bethe approximation.

\section*{Conclusion}

We have considered the effect of the assumption of neighboring correlations for a two-dimensional hardcore boson model on a square lattice within Bethe approximation using the clusters made from two and
four sites. 
Explicit equations for critical temperatures of charge and superfluid ordering within mentioned approximations have been derived. 
Analysis of the obtained solutions for various ratios of the parameter of inter-sites charge correlations and transfer integral $V/t$ has shown fundamental differences of the concentration dependences of critical temperatures as compared to MFA. 
A critical concentration of bosons, which restricts the existence domain of solutions of CO type, arises for the temperatures of charge ordering. 
In the case of superfluid ordering, there is a reduction of critical temperature up to zero values at half filling. 
Phase diagram of the hard-core boson model has been constructed, which is obtained within two- and four-site Bethe approximation at $V/t = 3$ with the assumption of phase separation within Maxwell's construction.
It has been shown that taking into account of short-range correlations within Bethe approximation approaches the form of phase diagram to the results obtained by quantum Monte Carlo~\cite{Schmid2002}.

\bigskip
The reseach was supported by the Government of the Russian Federation, Programm 02.A03.21.0006 and by the Ministry of Education and Science of the Russian Federation, project No FEUZ-2020-0054.


\end{document}